# MORPH: A Reference Architecture for Configuration and Behaviour Self-Adaptation

Victor Braberman*  Nicolas D'Ippolito*  Jeff Kramer†, Daniel Sykes†, and Sebastian Uchitel*†
† Department of Computing, Imperial College London, UK.
* CONICET and Departamento de Computación, FCEN, Universidad de Buenos Aires, Argentina.

*Abstract*—An architectural approach to self-adaptive systems involves runtime change of system configuration (i.e., the system's components, their bindings and operational parameters) and behaviour update (i.e., component orchestration). Thus, dynamic reconfiguration and discrete event control theory are at the heart of architectural adaptation. Although controlling configuration and behaviour at runtime has been discussed and applied to architectural adaptation, architectures for self-adaptive systems often compound these two aspects reducing the potential for adaptability. In this paper we propose a reference architecture that allows for coordinated yet transparent and independent adaptation of system configuration and behaviour.

## I. INTRODUCTION

Self-adaptive systems are capable of altering at runtime their behaviour in response to changes in their environment, capabilities and goals. Research and practice in the field has addressed challenges of designing these systems from multiple perspectives and levels of abstraction.

It is widely recognised that an architectural approach to achieve self adaptability promises a general coarse-grained framework that can be applied across many application domains, providing an abstract mechanism in which to define runtime adaptation that can scale to large and complex systems [1].

Architecture-based adaptation involves runtime change of system configuration (e.g., the system's components, their bindings, and operational parameters) and behaviour update (e.g., component orchestration).

Existing approaches to architectural adaptation (e.g. [2], [3], [4], [5], [6], [7], [8]) incorporate elements from two key areas to enable runtime adaptation: Dynamic reconfiguration [9], [10], [11], [12], [13], [14], [15], [1], [1] and discrete-event control theory [14], [16], [17], [18], [19], [20], [21], [22], [23], [24]. The first, key for adapting the system configuration, studies how to change component structure and operational parameters ensuring that on-going operation is not disrupted and/or non-functional aspects of the system are improved. The second, key for adapting behaviour, studies how to direct the behaviour of a system in order to ensure high-level (i.e., business, mission) goals.

Although the notions of configuration and behaviour control are discussed and applied by many authors, they are typically compounded when architectures for adaptation are presented, reducing overall architectural adaptability. Automated change of configuration and behaviour addresses different kinds of adaptation scenarios each of which should be managed as independently as possible from the other. Nonetheless, configuration and behaviour are related and it is not always possible to change one without changing the other. The need for both capabilities of independent yet coordinated adaptation of behaviour and configuration requires an extensible architectural framework that makes explicit how different kinds of adaptation occur.

*Consider a UAV on a mission to search for and analyse samples. A failure of its GPS component may trigger a reconfiguration aiming at providing a location triangulating over alternative sensor data. The strategy may involve passivating the navigation system, unloading the GPS component and loading components for other sensors in addition to the component that resolves the triangulation. A behaviour strategy that is keeping track of the mission status (e.g. tracking areas remaining to be traversed, samples collected, etc.) should be oblivious to this change.*

A reconfiguration adaptation strategy that can cope with the GPS failure can be computed automatically using approaches based on, for example, SMT solvers or planners [25], [16] that consider the structural constraints provided in the system specification (e.g., the need for a location service), requirements and capabilities of component types (e.g., the requirements of a triangulation service) and runtime information of available component instances (e.g., the availability of other sensors).

*The arrival of the UAV at an unexpected location due to, say, unanticipated weather conditions may make the current search and collection strategy inadequate. For instance, the new location may be further away from the base than expected and the remaining battery charge may be insufficient to allow visiting the remaining unsearched locations before returning to base. In this situation the behaviour strategy would have to be revised to relinquish the goal of searching the complete area before returning to base in favour of the safety requirement that battery levels never go below a given threshold. The new behaviour strategy may reprioritise remaining areas to be searched (in terms of importance and convenience), visiting only a subset of the remaining locations as it moves towards the base station for recharging. Once recharged, the strategy may opt to attempt to revisit the entire area under surveillance but prioritising the locations previously discarded. Such behavioural adaptation should be independent to the infrastructure supporting reconfiguration control.*

A behaviour strategy that can deal with unexpected deviations in the UAV's navigation plan can be computed auto-matically using approaches based on, for instance, controller synthesis [20] that consider a behaviour model describing the capabilities of the UAV (e.g. autonomy), environment (e.g., map with locations of interest and obstacles) and system goals (e.g. UAV safety requirements and search and analyse – liveness – requirements). Indeed, our proposal of an explicit separation of reconfiguration and behaviour strategy computation and enactment is in line with the design principles of a separation of concerns and information hiding [26]. The behaviour strategy is oblivious to the implementation that provides the services it calls, while the reconfiguration strategy supports the injection of the dependencies that are required by the behaviour strategy oblivious to the particular ordering of calls that the behaviour strategy will make. In a sense, the design principle which is known to support changeability supports runtime changeability, which ultimately is what adaptation is about.

Configuration and behaviour adaptation may however need to be executed in concert. *Consider the scenario in which the gripper of the UAV's arm that is to be used to pick up samples becomes unresponsive. With a broken gripper the original search and analyse mission is unachievable. This should trigger an adaptation to a degraded goal that aims to analyse samples via on-board sensors and remote processing. This goal requires a different behaviour strategy (e.g. circling samples once found to perform a 360 degree analysis) but also a different set of services provided by different components (e.g. infra-red camera). Not only are both behaviour and configuration adaptation required, but also their enactment requires a non-trivial degree of provisioning: To set up the infra-red camera, the UAV requires folding the arm to avoid obstructing the camera's view; performing such an operation while in the air is risky. Hence, coordination between configuration and behaviour adaptation is needed: First, a safe landing location must be found, then arm folding must be completed, and only then can the reconfiguration start. New components are loaded and activated, and finally, a strategy for in-situ analysis, rather than analysis at the base, can start.*

It is in the combined configuration and behaviour adaptation where the need for both separation of concerns and explicit architectural representation of coordination becomes most evident. Approaches to automated computation of configuration and behaviour adaptation strategies require different input information and utilise different reasoning techniques. Both automated reasoning forms are of significant computational complexity and require careful abstraction of elidable information. Keeping resolution of configuration and behaviour adaptation separately allows reuse of existing and future developments in the fields of dynamic reconfiguration and control theory and also helps keep computational complexity low.

*The broken UAV gripper scenario requires a coordinated behaviour and reconfiguration adaptation strategy. The adaptation required can be decomposed into a behaviour control problem that assumes that a reconfiguration service is available and a reconfiguration problem. The resulting behaviour strategy will be computed on the assumption that the UAV's capabilities will conform to the current configuration (e.g. grip command fails) until a reconfigure command is executed, and that from then on different capabilities will be available (e.g. infra-red camera getPicture command available). The behaviour strategy computation will also consider restrictions on when the reconfigure command is allowed (e.g. when arm is folded) and new goals (360 degree picture analysis rather than collect). The computation of the reconfiguration strategy does not entail additional complexity and is oblivious to the fact that a behaviour strategy that involves a reconfiguration halfway through is being computed.*

In the above scenario, what needs to be resolved at the architectural level of the self-adaptation infrastructure is which architectural element is responsible for the decomposition of the adaptation strategy into a behaviour strategy and a reconfiguration strategy, and also how strategy enactment is performed to allow the behaviour strategy to command reconfiguration at an appropriate time (and possibly even account for reconfiguration failure). Indeed, an appropriate architectural solution to this would enable guaranteeing that given a correct decomposition of the overall composite adaptation problem into configuration and behaviour adaptation problems, and given correct-by-construction configuration and behavioural strategies for these problems, the overall adaptation problem is correct.

In this paper we present MORPH, a reference architecture for behaviour and configuration self-adaptation. MORPH makes the distinction between dynamic reconfiguration and behaviour adaptation explicit by putting them as first class entities. Thus, MORPH allows both independent reconfiguration and behaviour adaptation building on the extensive work developed but also allowing coordinated configuration and behavioural adaptation to accommodate for complex self-adaptation scenarios.

## II. MORPH: A Reference Architecture for Configuration and Behaviour Self-Adaptation

We start with a very brief introduction of the main architectural elements to give a general picture of how the architecture works before we go into detail of the workings and rationale of each element. A graphical representation of the architecture can be found in Figure 1. In the remaining text, when we want to emphasise traceability to the figure we will use an alternative font.

The architecture is structured in three main layers that sit above the target system: Goal Management, Strategy Management and Strategy Enactment. Orthogonal to the three layers is the Common Knowledge Repository. Each layer can be thought of as a implementing a MAPE-K loop. The top layer's MAPE-K loop is responsible for reacting to changes in the goal model that require complex computation of strategic, possibly configuration and behavioural, adaptation. Its knowledge base is the Common Knowledge Repository. The Strategy Management layer's MAPE-K loop is responsible for adapting to changes that can be addressed using pre-processed strategies. It selects pre-computed strategies based on the Common Repository Knowledge and a set of internally managed pre-computed strategies. The Strategy Enactment layer's MAPE-K loop is responsible for executing strategies; its knowledge base is primarily the strategy under enactment.

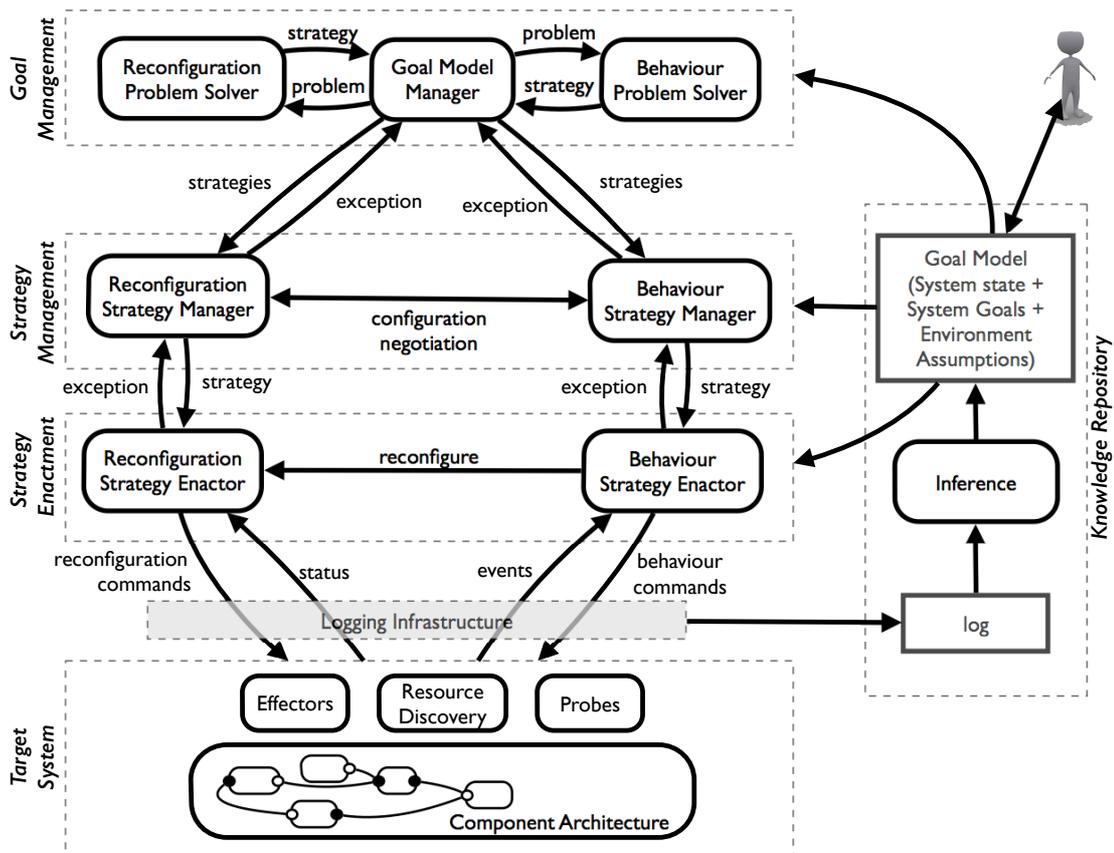

Fig. 1. The MORPH Reference Architecture.

The Target System abstracts the Component Architecture that provides system functionality. The Component Architecture is harnessed by effectors and probes which allow the Strategy Enactment Layer to interface with system components. The Knowledge Repository stores in a Log the execution data produced by Target System and also stores in the Goal Model) the result of Inference procedures that produce knowledge regarding the system state, goals and environment assumptions. We expect users, administrators and other stakeholders to also produce modifications to the Knowledge Repository, and in particular the goals and environment assumptions.

The three layers that provide the architectural adaptation infrastructure are each split into reconfiguration and a behaviour aspects. The Goal Management layer has a Goal Model Manager whose main responsibility is to decompose adaptation problems into reconfiguration and behaviour problems, each of which is given to a specific Solver to produce a strategy that can achieve the required adaptation. The top layer sends reconfiguration and behaviour strategies downwards. The bottom two layers have architectural elements to handle reconfiguration and behaviour strategies separately but interact with each other when and if needed to maintain overall consistency. The Strategy Management layer entities interact to ensure that they select consistent strategies to be executed by the Strategy Enactment layer (c.f., configuration negotiation). The Strategy Enactment layer entities interact to ensure that the execution of their respective strategies is done consistently over time (c.f., reconfigure command).

### A. Target System

**Responsibility**: The Target System's responsibility is to achieve the system goals, encapsulate implementation details and provide abstract monitoring and control mechanisms over which the behaviour and structure of the system can be adapted.

**Rationale**: The rationale for this subsystem is to encapsulate the instrumentation of the system-to-be-adapted to support a flexible and reusable framework for monitoring, analysing, planning and executing adaptation strategies in the layers above.

**Structure and Behaviour**: The Target System, strongly inspired by [3], contains the component architecture that provides the managed system's functionality *(e.g., GPS, video, telemetry and navigation components)*. It also contains instrumentation to monitoring and control of the component architecture. Two types of effectors are provided. The first provides an API to add, remove and bind components, in addition to setting operational parameters of these components. We refer to the invocation of operations on these effectors as reconfiguration commands. These effectors are application domain independent, and they provide the adaptation infrastructure an abstraction over the concrete deployment infrastructure on which the component architecture runs *(e.g., the UAVs operating system)*. The second effector type, behaviour actions, is domain dependent and provides an API that invokes functional services provided by components of the component architecture. *The UAV's navigation component may exhibit a complex API which is abstracted into simple commands (e.g. goto(Location)) that are to be used as the basis for behaviour strategies.*

The mechanism for monitoring of the component architecture can be provided by probes that reveal state information. As with effectors, monitoring information can be classified into two kinds. We have on one hand information regarding the status of components. This kind of information is application independent. Status of a component may indicate if it is active, inactive, connected or killed as in [25] or indicate its modes of operation as in [27]. On the other hand we refer to as events the application domain relevant information that flows from the Target System to the Strategy Enactment Layer. *UAV events may include notifications regarding battery depletion, or acknowledgements of having reached a requested location.*

Between the target system and the adaptation infrastructure a translation layer is required to provide translation services that aim to bridge the abstraction gap between the knowledge representation required to perform adaptation at the architectural level and the concrete information of the actual implementation. *In the UAV, this may include resolving event handlers, process ids, in addition to domain specific translations such the conversion of continuous variable for battery level to a discrete battery depleted event.*

### B. Common Knowledge Repository

**Responsibility**: The key responsibility of the repository is to keep an up to date goal model at runtime based on inferences made over continuous monitoring of the environment to detect changes in goals, behaviour assumptions and available infrastructure.

**Rationale**: The design rationale for the repository is to decouple the accumulation of runtime information of the target system from the complex computational processes involved in abstracting and inferring high-level knowledge that can be incorporated, for subsequent adaptation, into a structured body of knowledge regarding stakeholder goals, environmental assumptions and target system capabilities.

**Structure and Behaviour**: The common knowledge repository stores information about the target system, the system goals and environment assumptions. It consists of two data structures (a log and a goal model) and an Inference procedure.

Information about the target system refers to historical and current information about the behaviour of the system being adapted by the three-tier adaptation infrastructure. This information includes a low level log of the evolution of the state of system components (*e.g., minute by minute recording of battery level*) and messages between them (*e.g., GPS location request by navigation component*) to aggregate data, possibly computed through complex inference procedures, including statistical information related to reliability or performance (*e.g., battery consumption ratio and predicted depletion time*).

The Goal Model: This is the key data architectural element of the repository. We use the term "goal model" in the sense of goal oriented requirements engineering [28], [29] inspired on the world-machine model for requirements engineering [30], [31] where the focus is on structuring information about the purpose of the socio-technical system and how such objectives can be achieved in different ways based on combinations of assumptions on the environment and requirements on the software. The purpose is sometimes referred to as high-level or business goals.

Goal models structure goals (state-based assertions intended to be satisfied over time) into refinement structures. An AND-refinement structure describes how goals can be achieved by achieving all their subgoals (*e.g., search and analyse can be acheived discretising the area into 100 square meter regions, visiting each one, using image recongition to identify samples, fine grained navigation to land next to them and an arm to pick them up and take them to base*). An OR-refinement structure encodes alternative ways in which a goal may be achieved (*e.g., rather than picking samples and taking them to base, they can be analysed in-situ with an infra-red camera*). The refinement structures define an acyclic directed graph where leaf nodes need to be assigned to an entity that will guarantee them. These entities can part of the software (i.e., components) or the environment (i.e., users, external systems or laws of nature). Nodes assigned to the environment form domain assumptions (*e.g., the shape and density of samples is such that the gripper will firmly hold them and allow them to be picked up*). Goals assigned to software components are requirements (*e.g., commanding the gripper to close will make fingers clasp small objects within a 5cm radius of its palm*).

If an AND-refinement of a top level goal is correct and all leaf nodes of the refinement are either valid assumptions or requirements guaranteed by the components they are assigned to, then the top level goal is guaranteed. OR-refinements represent alternative strategies (AND-refinements) for achieving top level goals. The selection of a specific or-refinement can depend on soft goals indicating stakeholder preferences. These preferences can be based, for instance, on non-functional requirements (*e.g., preferring a more precise location service for the UAV, preferring sample analysis to be conducted at base rather than in-situ*). Soft goals also form part of the goal model and are amenable to be structured into refinement structures.

The point of keeping a structured view of the world that includes requirements and assumptions, with multiple ways of achieving high-level goals and preference criteria over these alternatives is that at runtime it is possible to change the way a goal is achieved by selecting a different OR-refinement. The combinatorial explosion of possible OR-refinement resolutions can be a rich source for adaptation which is exploited in the Goal Management Layer. In addition, this representation of rationale is amenable to being updated and changed automatically as new information is acquired.

Note that we do not prescribe a particular representation of the domain knowledge for this repository. For instance, architectural description languages (e.g., [32], [33]) can be used for expressing structural requirements, tabular formats (e.g., [34]) or contract-based specifications for describing behaviour requirements, and automata based languages (e.g., [35]) for expressing environment behaviour assumptions. In this paper, we have used the terminology of goal oriented requirements engineering as a way of conceptualising the information about assumptions, requirements and system objectives and state that is necessary to reason about self-adaptation.

### C. Goal Management Layer

**Responsibility**: The main responsibility of the Goal Management Layer is to deal with and anticipate changes in the stakeholder goals, environment assumptions and system capabilities by pre-computing adaptation strategies consisting of separate behaviour and reconfiguration strategies.

**Rationale**: The rationale for this layer is based on two core concepts: The first is that the adaptive system must be capable of performing strategic, computationally expensive, planning independently of and concurrently with the execution of pre-computed strategies (occurring in lower layers). The second is to decompose adaptation into a behaviour strategy that controls the system to an interface and a reconfiguration strategy that injects the dependencies on concrete implementations that the behaviour strategy will use. Decomposing adaptation along the modular design improves support for adaptability allowing behaviour and configuration changes independently.

**Structure and Behaviour**: The layer has three main entities, the Goal Model Manager, the Behaviour Problem Solver and the Reconfiguration Problem Solver.

The Goal Model Manager: This is the key element of the layer, and is responsible for three core tasks: The first is to decide when a new adaptation strategy must be computed, the second is to resolve all OR-refinements in the goal model and select the requirements to be achieved by the system, and third is to decompose the requirements into achievable reconfiguration and behaviour problems. The concrete strategies for reconfiguration and behaviour are computed by the solvers.

Production of adaptation strategies can be triggered by requests for plans from layers below or internally due to the identification of significant changes in the goal model. The former case corresponds to a scenario in which a failure is propagated rapidly upwards from the target system: *For instance, the UAVs gripper component fails. The Strategy Enactment layer, which is executing a strategy that requires the gripper, immediately declares that its current strategy is unviable and requests a new strategy from the Strategy Management Layer. However, all pre-computed plans in the middle layer are based on having some form of arm to pick up objects to be studied. Consequently, the computation of a new strategy for achieving system goals is requested to the Goal Model Manager.*

The alternative, internal, triggering mechanism corresponds to scenarios in which the goal model is changed because of new information inferred from the log or input manually by some stakeholder. *For instance, weather conditions may lead to inferring higher energy consumption rates from logged information. What would follow is a revision of the assumptions on UAV autonomy stored in the Goal Model. Such alteration may trigger the re-computation and downstream propagation of search strategies to make more frequent recharging stops.*

The selection based on soft goals of a specific set of requirements (leaf nodes assigned to the target system), resulting from a particular or-refinement resolution, requires information about the current system state and possibly aggregate information on its past execution (*as with energy consumption*). For instance, with no components capable of picking up samples for transportation, an OR-refinement representing degraded levels of service (in which samples are inspected in-situ rather than at base) will only be viable. On the other hand, with a gripper component functioning, a preference on the quality of sample inspection will lead to selecting strategies that perform base-located inspections. Another example may be the selection of the position system used (GPS vs. hybrid positioning) based on component availability and precision.

Adaptation strategies are decomposed into a strategy for achieving the component configuration that can provide the functional services required to achieve selected requirements and a behaviour strategy that can call these services in an appropriate temporal order to satisfy the requirements. *The adaptation strategy that deals with the broken gripper must reconfigure the system to use a different set of components (e.g. the infra-red camera) and coordinate its use upon reaching a position where there is a sample to be inspected.*

As discussed in the Introduction section, decomposition allows adaptation of the system configuration transparently to the behaviour strategy being executed (*e.g., changing the location mechanism*) or the behaviour strategy transparently to the configuration in use (*e.g., changing the route planning strategy*). In addition, decomposition allows the computation of multiple behaviour strategies for a given configuration (*e.g. different search and collect strategies that assume different UAV autonomy can be run on a configuration that has a gripper component*) and different configurations can be used for a given behaviour strategy (*e.g. different configurations for providing a positioning service can be used for the same search strategy*).

One of the design rationales for this layer is the pre-computation of expensive adaptation strategies that are then ready to use when needed. This means that multiple reconfig-uration and behaviour strategies may be constructed. Indeed, the Goal Model Manager can pre-compute, and propagate downwards, many reconfiguration and behaviour strategies for one resolution of the OR-refinements of the goal model. *This may be useful, for example, if it is known that information regarding UAV autonomy is imprecise, multiple (behaviour) search strategies for searching the area may be developed so that the infrastructure can adapt quickly as soon as the predicted UAV autonomy differs significantly from what can be inferred from the monitored energy consumption. Similarly, should the GPS-based location service be known to fail (perhaps do to environmental conditions), then various reconfiguration plans may be pre-computed to allow adaptation to alternative positioning systems when needed.* Furthermore, adaptation strategies for different resolutions of the goal models OR-refinements may be pre-computed. *For instance, upon detecting an unreasonably high failure rate of the gripper component attempting to pick samples up, the goal model manager may produce a strategy for a degraded system objective (inspect in-situ) that does not require a gripper component. Should the gripper finally fail completely (or its failure rate become unacceptably low) then the pre-computed strategy can be put in place rapidly.*

Pre-computation of multiple reconfiguration and behaviour strategies for different OR-refinement resolutions requires keeping additional consistency information. A many to many consistency relation must be maintained by the data structures representing reconfiguration and behaviour strategies so to allow lower levels to ensure consistent selection of a strategy of each kind. Furthermore, a relation between these strategies and the OR-refinement resolutions that they have been designed for must also be kept and propagated downstream to lower layers.

<u>Configuration Problem Solver</u>: The layer has two entities capable of automatically constructing strategies for given adaptation problems. The Configuration Problem Solver focuses on how to control the target system to achieve a specified configuration given the current system configuration, configuration invariants that must be preserved and component availability. The configuration to be reached may be a partially specified. *For instance, the target configuration may be required to have a component that provides a positioning service. Ensuring that a GPS component is instantiated in the target system could satisfy such a requirement. However, a component using a hybrid positioning system may require additional components to be present (e.g. WI-FI, Bluetooth, Mobile phone) and bound to it.* Configuration invariants may include structural restrictions forcing the architecture to conform to some architectural style or other considerations based, for instance, on non-functional requirements. *In the UAV such restrictions may include that the attitude control components never be disabled or that the total number of active components never be beyond a given threshold to avoid battery overconsumption.*

Reconfiguration problem solvers build strategies that call actions that add and remove components, activate and passivate them, and establish or destroy bindings between them. These reconfiguration commands are part of the API exposed by the target system. The strategy may sequence these actions or have an elaborate scheme that decides which actions to call depending on feedback obtained through the information on the status of components exhibited by the target system API.

To automatically construct strategies the solvers can build upon a large body of work developed in the Artificial Intelligence and Verification communities, including automatic planners (e.g. [36]), controller synthesis (e.g. [21]), and model checking (e.g. [37]). Such techniques have been applied to construction of reconfiguration strategies in [7], [25], [16]

<u>Behaviour Problem Solver</u>: This entity focuses on how to control the target system to satisfy a behaviour goal. In contrast to reconfiguration problems, the behaviour goal may not be restricted to safety and reachability (i.e. reach a specific global state while preserving some invariant). Behaviour goals may include complex liveness goals such as to have the UAV monitor indefinitely an area for samples to inspect. Behaviour problem solvers produce strategies, which can be encoded as automata that monitor target system events and invoke target system actions.

In addition to the expressiveness of goals that behaviour strategies must resolve, there is an asymmetry between reconfiguration and behaviour problems. To resolve the coordination problem between strategies (*as with folding the UAV arm before a reconfiguration to deal with a gripper failure can be executed, see Introduction*), the behaviour strategies produced by the solver can invoke a reconfigure command, which triggers the execution of a reconfiguration strategy (see Section II-D).

Techniques that build automatically behaviour strategies are typically based on planning and controller synthesis techniques and have been used in approaches such as [17], [38], [39].

*D. Strategy Management Layer*

**Responsibility**: This layer's main responsibility is to select and propagate pre-computed behaviour and reconfiguration strategies to be enacted in the layer below. For this, the layer must store and manage pre-computed behaviour and reconfiguration plans, and request new strategies to the layer above when needed. It must also ensure that the behaviour and reconfiguration strategies sent to the lower layer are consistent, indicating their relationships.

**Rationale**: They main concept for the layer is to allow rapid adaptation to failed strategy executions (or capitalising rapidly on opportunities offered by new environmental conditions) by having a restricted universe of pre-computed alternative behaviour and reconfiguration strategies that can be deployed independently or in a coordinated fashion.

**Structure and Behaviour** The layer has two entities that work in similar fashion mimicking much of the layer's responsibilities but only on either behaviour or reconfiguration strategies. However, the Behaviour Strategy Manager and the Reconfiguration Strategy Manager are not strictly peers. In some adaptation scenarios the former will take a Master role in a Master-Slave decision pattern.

Behaviour Strategy Manager: The manager stores multiple behaviour strategies from which it picks one to be enacted in the layer below. The selection of strategy may be triggered by an exception raised by the layer below or internally due to a change identified in the common knowledge repository. The former may occur when the behaviour strategy being executed finds itself in a unexpected situation it cannot handle. *For instance, the UAV executing a particular search strategy expects to be at a specific location with at least 50% of its battery remaining but finds that it is below that threshold, invalidating the rest of the strategy for covering the area to be searched.* At this point the Strategy Enactment Layer signals that the assumptions for its current strategy are invalid and requests a new strategy to this layer.

The other scenario that can trigger the selection of a new strategy is a change in the common knowledge repository. *Consider again the problem of unexpected energy overconsumption. An inference process in the knowledge repository may update the average energy consumption rate periodically based on Target System information being logged. This average may be well above the assumed consumption average for the behaviour strategy being executed. The Behaviour Strategy Manager may decide that it is plausible that the current behaviour strategy will fail and may decide to deploy a more conservative search strategy.*

Note that the two channels that may trigger the selection of a new strategy differ significantly in terms of latency and urgency. The exception mechanism provides a fast propagation channel of failures upwards, indicating that the strategy being currently enacted is relying on assumptions that have just been violated. This means that any guarantees on the success of the current strategy in satisfying its requirements are void and a new strategy is urgently required. The second channel is via de knowledge repository. The monitoring of changes in the knowledge repository is a process that incurs comparatively significant delays as the inference of goal model updates based on logged information may be performed sporadically and consume a significant amount of time. The upside of this second channel is that it may predict problems sufficiently ahead of their occurrence, providing time to select pre-computed strategies that may avoid them.

The selection of a behaviour strategy is constrained by the current configuration of the target system (which determines the events and actions that can be used by the strategy) and the alternative configurations that may be reached by enacting one of the pre-computed re-configuration strategies. Furthermore, the selection is informed by preferences defined in the goal model on which OR-refinement resolution is preferred. *Thus, a new search strategy that can be supported by the current UAV configuration may be selected. Alternatively a strategy that no longer picks samples up to avoid the extra consumption produced by load carrying may be chosen. In the later case, in-situ analysis is required and hence a reconfigured UAV with an infra-red camera in place is required. Selecting such a pre-computed behaviour strategy is subject to the availability of a pre-computed reconfiguration strategy that can reach a configuration with an active infra-red camera module.*

The Behaviour Strategy Manager deploys the selected strategy by performing two operations. Firstly, should the selected strategy require a configuration with characteristics that are currently not provided, it commands the Reconfiguration Strategy Manager to deploy an appropriate reconfiguration strategy (c.f. Master-Slave relationship). Secondly, the manager hot-swaps the current behaviour strategy being executed in the layer below with the newly selected strategy, setting the initial state of the new strategy consistently with the current state of the Target System. Note that should the new strategy be replacing a strategy that is still valid (i.e. no exception has been raised) then the hot-swap procedure may also exploit information extracted from the current state of the strategy to be swapped out.

Should the Behaviour Strategy Manager fail to select a pre-computed behaviour strategy, the manager requests new strategies from the layer above. This may happen, for example, because none of pre-computed strategies it manages have assumptions that are compatible with the actual observed behaviour of the system (e.g., *energy consumption is far worse than what is assumed by any pre-computed strategy*) or that they all rely on unachievable configurations (e.g. *the joint failure of the gripper component and infra-red camera was a operational scenario not considered in any of the pre-computed strategies*).

Reconfiguration Strategy Manager: This entity works similarly to its behaviour counterpart. It stores and manages multiple reconfiguration strategies and selects them for deployment constrained by the availability of components in the Target System while maintaining consistency with the configuration requirements of the current behaviour strategy. Selection is also informed by preferences specified in the goal model. *Consequently, a precision preference may lead to selecting a reconfiguration strategy that attempts to use a GPS rather than a hybrid positioning component.*

When negotiating with the Behaviour Strategy Manager on a pair of strategies to be deployed, the Reconfiguration Strategy Manager takes the slave role, stating the configurations requirements that are achievable and then selecting an appropriate reconfiguration strategy based on the selection made by the Behaviour Strategy Manger.

There are three channels that can trigger the selection of a new reconfiguration strategy. Two are similar to those that trigger the Behaviour Strategy Manager: An exception from the Reconfiguration Strategy Enactor and a change in the goal model. *Examples of these are the failure of the GPS component triggering a rapid response by the manager which selects an alternative configuration (using the hybrid positioning component) and deploys an appropriate reconfiguration strategy, or an increased response time of the GPS component leading to the decision of changing the positioning system before it (most likely) fails.* The third channel is the request of a new configuration by the Behaviour Strategy Manager (which in turn may have been triggered via the exception mechanism or a change in the goal model).

It is important to note that deployment of new strategies at the Strategy Management layer may respond not only to problems (or foreseen problems) while enacting the current strategies, but also to capitalise on opportunities afforded by a change in the environment. For instance, should a new component become available, or statistics on its performance improve, this would be reflected in the knowledge repository and an alternative preferred pre-computed strategy may be deployed.

*E. Strategy Enactor*

**Responsibility**: This layer's main responsibility is to execute behaviour and reconfiguration strategies provided by the layer above. Strategy execution involves monitoring the target system and invoking operations on it at appropriate times as defined by the strategy. The layer must also ensure that if the target system reaches a state unexpected by the strategy, this should be reported to the layer above.

**Rationale**: The aim is to provide a MAPE loop with low analysis latency to allow rapid response to changes in the state of the target system based on pre-computed strategies. In other words to achieve fast adaptation to anticipated behaviour of the target system. IN addition, to allow independent handling of failed assumptions made by either the behaviour or reconfiguration strategies, thereby adapting one strategy in a way that is transparent to the other.

**Structure and Behaviour**: The layer has two strategy enactors, one for behaviour strategies and the other for reconfiguration strategies. Both enactors work very similarly. They monitor the target system and react to changes in the system by invoking commands on the target system. The decision of which command to execute requires no significant computation. The two enactors do, however, differ in the instrumentation infrastructure they use to monitor and effect the target.

<u>Reconfiguration Strategy Enactor</u>: This entity invokes reconfiguration commands and accesses individual software component status information through an API provided by the Target System layer. The aspects monitored and effected by this enactor tend to be application domain independent; commands and status data are typically related to the component deployment infrastructure and allow operations such as adding, removing and binding components, and checking if they are idle, active, and so on. Commands can include setting operational parameters of components (eg., thread pool).

In addition to sequencing reconfiguration commands, the enactor has to resolve the challenge of ensuring that state information is not lost when the configuration is modified. This can involve ensuring stable conditions such as tranquility [13] or quiescence [12] before change. For instance, *the infra-red camera component can be safely removed from a system if it is isolated (no bindings to or from) and passive (e.g., not processing an image)*.

<u>Behaviour Strategy Enactor</u>: The entity monitors and affects the target system through application domain services provided by the components of the target system. These are accessed via behaviour commands and event abstractions exhibited by the Target System layer. The enactor starts executing the behaviour strategy assuming that there is a configuration in place that can provide the events and commands it requires. *Thus, a new search and analyse behaviour strategy using the gripper is assuming the gripper component is configured.*

Should the behaviour strategy require a different configuration at any point, it must request the configuration change explicitly. In this case a reconfigure command will be part of the behaviour strategy and the behaviour enactor will command the execution of the reconfiguration strategy stored by the reconfiguration strategy enactor (*e.g., the behaviour strategy folds the arm holding the broken gripper and then requests reconfiguration to incorporate the infra-red camera to only then proceed with in-situ analysis*). Note that in this case the behaviour enactor assumes that the reconfiguration strategy loaded in the other enactor will attempt to reach a target configuration that is consistent with the behaviour strategy.

Assumptions regarding the current configuration and the target configuration of the strategy loaded on the reconfiguration strategy enactor are ensured by the layer above that feeds consistent behaviour and reconfiguration strategies to this layer.

## III. RELATION TO EXISTING SYSTEMS AND ARCHITECTURES

*A. RAINBOW [3]*

Rainbow instantiates and refines the MAPE-K architecture providing an extensible framework for sensors and actuators at the interface between the control infrastructure and the target system (see Figure 2). The architecture recognises the complexity of the interface between the MAPE-K infrastructure (referred to by the authors as the architecture layer) and the component system to be adapted (referred to as the system layer). The Rainbow framework introduces additional infrastructure into the architecture and system layers in addition to accounting for an extra layer between the two: the translation layer. Monitoring is split amongst the three layers: probes are introduced as system layer infrastructure to support observation and measurement of low-level system states. Gauges are part of the architectural infrastructure layer and aggregate information from the probes to update appropriate properties of the knowledge base used for the MAPE activities. The translation layer resolves the abstraction gap between the system layer and the architectural layer, for instance relating abstract component identifiers in the later concrete process and machine identifiers in the former.

Rainbow focuses on achieving self-adaptation through configuration adaptation. Thus, focus is on changing component instances and bindings and also effecting behaviour by changing operational parameters (thread pool size, number of servers, etc.). Indeed, the framework does not account explicitly for automated construction of strategies that control the functional behaviour of the system layer components. Thus, the distinction and coordination between configuration and behaviour control is not elaborated explicitly in the architecture.

MORPH takes inspiration from Rainbow when structuring the instrumentation of the managed system, thus including a translation layer, effector and probes. It also recognises the complexity of the data accumulation and analysis aspects which in Rainbow are set in an architectural element named Model Manager, while we place it in the Knowledge Repository. The latter serves as the "K" element for multiple MAPE-K loops in MORPH, while in Rainbow the Model Manager services the only MAPE-K loop under execution.

Rainbow's adaptation engine requires a set of precomputed strategies and tactics, and in this sense, it can be said that the MAPE-K loop it implements falls within the Strategy Management layer. Computationally complex construction of strategies is not considered in Rainbow (strategies are assumed to be provided), hence no counterpart to MORPH's Goal Management layer exists. Indeed, according to the MORPH reference architecture, the input to Raibow's Adaptation Engine (the strategies) would be provided by the Goal Management Layer (see Figure 2).

### B. PLASMA [16]

PLASMA is a three layered architecture (see Figure 3) supporting model based adaptation using planning as the core technology for producing adaptation strategies. The architecture supports both reconfiguration and behaviour adaptation. The former is achieved through *adaptation plans* while the latter through *application plans*. A key design concern of PLASMA is that adaptation infrastructure be adaptable itself, thus each layer is instantiated by the layer above. This is an important difference with MORPH in which such concern is not considered the primary driver. Nonetheless, their are important similarities between PLASMA and MORPH.

In the top PLASMA layer, the panning layer, sit two planners, the Application and the Adaptation Planners. These correspond to the behaviour and reconfiguration problem solvers of MORPH. However, in the top PLASMA layer there is no Goal Model Manager that decomposes the system goals into two. Rather, it is assumed that the system goal is first addressed as a behaviour problem and then an adequate reconfiguration is produced for the behaviour strategy that is computed. In a sense, a simplified Goal Model Manager is hard-wired into the layer. This is a key difference with the approach presented herein. In MORPH, the achievable configurations may determine the system goal that can be satisfied. Furthermore, to decouple computationally intensive problem solving from the rest of the system adaptation and execution mechanisms, MORPH allows computing multiple behaviour and reconfiguration strategies, maintaining a many to many relationship between them. This allows having multiple precomputed adaptation strategy pairs that can be deployed immediately when required.

The middle PLASMA layer is the adaptation layer. Its core element is the Adaptation Analyzer which executes reconfiguration strategies produced by the level above. Thus, the Adaptation Analyzer fits well with MORPH's Reconfiguration Strategy Enactor.

The bottom PLASMA layer is the application layer which contains elements that correspond to what here we call the Target System (the application's component architecture, effectors and probes) and also contains the Application Executor. The Executor enacts behaviour strategies, consequently mapping to MORPH's Behavior Strategy Enactor.

The two bottom PLASMA layers determine a dependency that is not present in MORPH. In PLASMA, it is the reconfiguration adaptation that monitors and commands the behaviour adaptation. We place the behaviour and reconfiguration on equal grounds in all layers and in particular in the Strategy Enactment layer were both enactors are. The dependency introduced in PLASMA entails that the reconfiguration strategy has to put in place the behaviour strategy, forcing a reconfiguration every time a new behaviour strategy is computed.

As discussed previously, to support adaptation independent yet coordinated behaviour and reconfiguration is required. In PLASMA, coordination is achieved by the hard-wired dependency between plan computation and the hierarchical precedence of reconfiguration over behaviour plan enactment. The cost of this coordination is the lack of independent configuration and behaviour adaptation. In MORPH, treating reconfiguration and behaviour planning and enactment as peers supports their independence, coordination is achieved by introducing a Goal Model Manager in the top layer, negotiation in the middle layer and only in the bottom layer, once strategies are guaranteed to be consistent, a temporal coordination dictated by the behaviour strategy.

### C. Three Layer Conceptual Model [2]

The need to deal with hierarchies of control loops in autonomous systems is widely recognised (e.g. [40]). Lower levels are typically low latency loops that focus on more tactic and stateless objectives that involve less monitored and controlled elements while higher levels tend to focus on more stateful and strategic objectives involving multiple controlled and monitored aspects that require higher latency loops.

The need for hierarchy in architectural self-adaptation is discussed in [2]. A three layer architecture is proposed to provide a separation of concerns and to address a key architectural concern related to dealing with the complexity of run-time construction of adaptation strategies. The architecture structures hierarchically the MAPE-K loops introducing a separation of concerns in which complex, strategic, resource consuming analysis is performed in the top layer (the Goal Management Layer), these plans are managed by the Change Management layer and enacted in the Component Layer.

Although the architecture is conceptual in nature, it does prescribe the kind of control that is effected on the adaptive system by establishing a clear interface between the adaptation infrastructure and the component based system to be adapted. The architecture assumes an interface on which it can take action on the current system configuration by creating and deleting components, binding and unbinding components through their required and provided ports and setting component modes (i.e., configuration parameters). Such interface is used for reconfiguration adaptation. Various instances of this architectural model have been implemented.

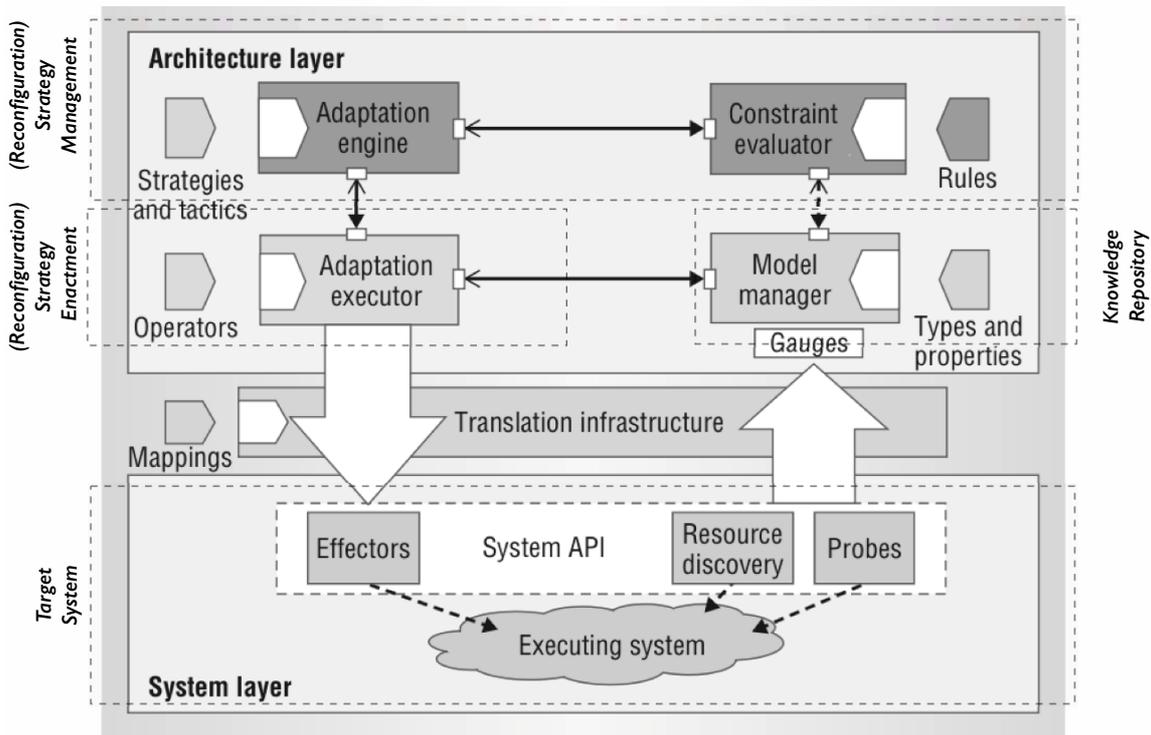

Fig. 2. The Rainbow Framework.

In [39], the three layer reference model is used to also adapt system behaviour. In this case planning is used to produce behaviour strategies in the Goal Management layer. The planner works on automatically inferred behaviour models of the environment. Each plan defines an interface which then must be matched with an appropriate configuration that can provide such interface. Thus a hierarchical relation, as in PLASMA (see above) is established between reconfiguration and behaviour adaptation, which as discussed previously hinders independent and coordinated behaviour and reconfiguration adaptation.

The MORPH reference architecture builds upon this three layered model emphasising the need to make behaviour and reconfiguration control first-class architectural entities and structuring how each works independently and in coordination. Thus, it provides a more refined view of the layers (as depicted in Figure 4).

## IV. PRIOR EXPERIENCE

The MORPH reference architecture systematically articulates our previous experience in concrete architectures and techniques for self-adaptation. This covers to different degrees, all elements in the reference architecture proposed herein (see Figure 5).

We describe the implementation of the Strategy Enactment Layer and how we instrument the Target System to support the Behaviour Strategy Enactor and the Reconfiguration Strategies Enactor in [41] and [1] respectively. In both cases, an interpreter is used to walk through a strategy executing command proxies that are bound to specific application components.

For the enactment of behaviour strategies [41], as the components are normally provided by third parties such as the robot arm manufacturer, each behaviour command may map to an ad-hoc combination of low-level method calls. The adaptive system designer must provide implementations of each high-level command (or event) that will be used in the behaviour strategy. This transformation of high-level behaviour commands into low-level method calls lives in the translation infrastructure that lies between the Target System and Strategy Enactment layer (not depicted in the MORPH architecture diagram) just as in [3] (see Figure 2)

The enactment of reconfiguration commands is done in [1] via the Backbone language [42], which is UML 2.0 compatible and resembles Darwin [32]. Thus high-level reconfiguration commands in the reconfiguration strategy are sent via RMI to the Backbone interpreter sitting in the target system (running on a Java VM) which then affects the component architecture appropriately. More details on Backbone can be found in [43].

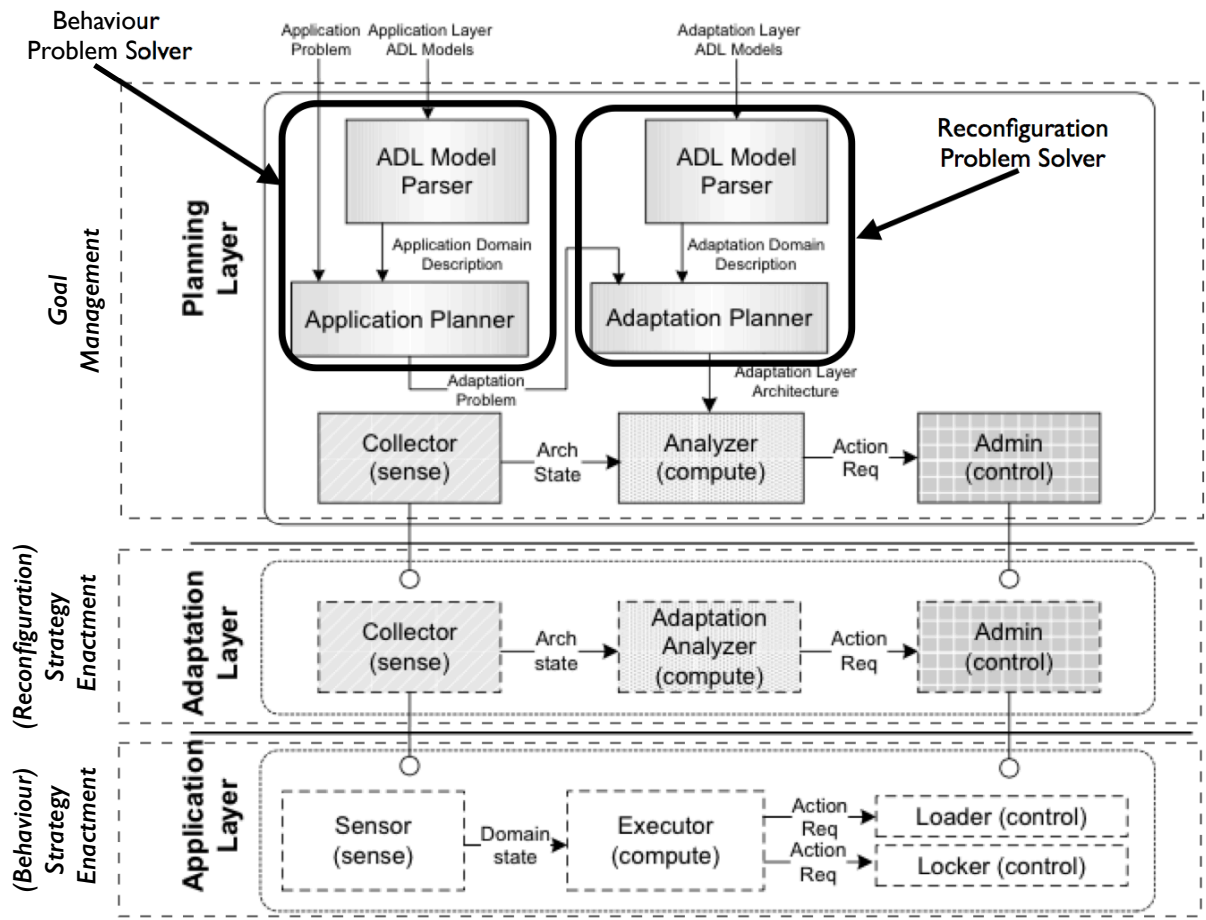

Fig. 3. The PLASMA Architecture.

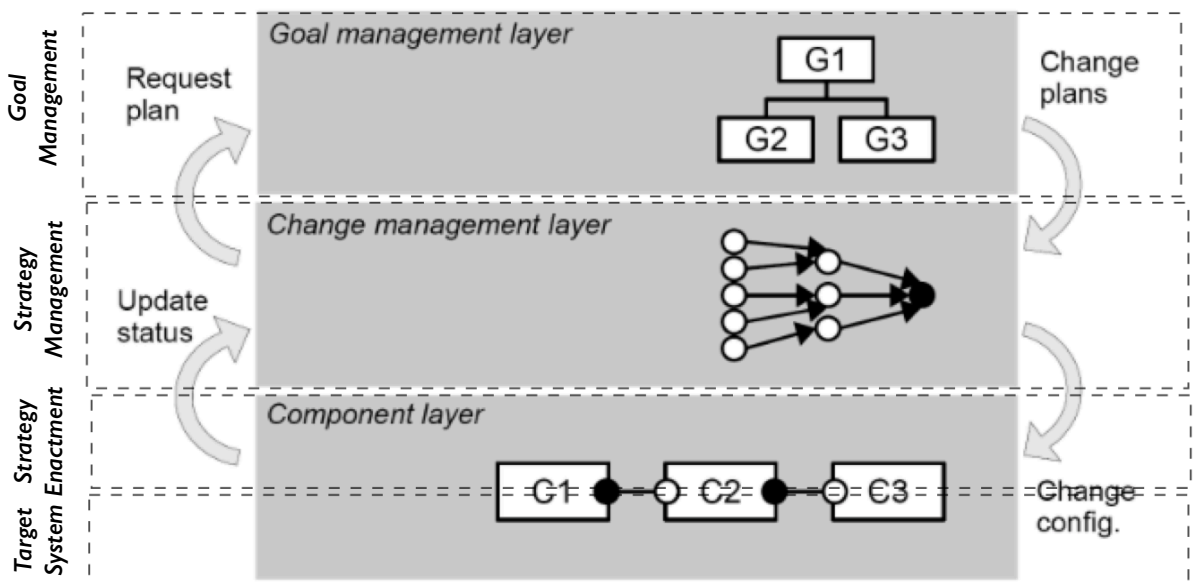

Fig. 4. The Three-Layer Conceptual Model [2]

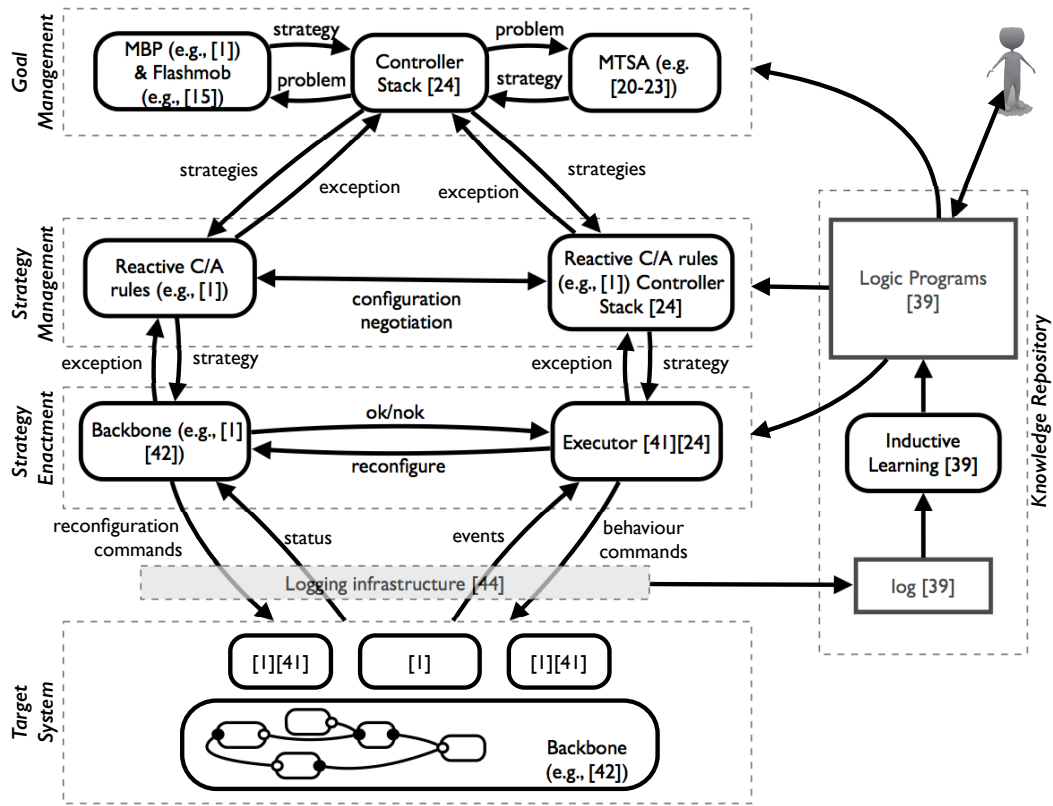

Fig. 5. Prior Experience.

In [1], we show how the Reconfiguration and Behaviour Strategy Managers can be equipped with general pre-computed strategies that can provide additional robustness to strategies running on a component based systems in which unexpected failures occur. The use of reactive condition-action rules allows re-sensing at regular intervals what the state of the system is to then enact an appropriate rule even if the effect of previously enacted rules was not as expected.

In [24], a more sophisticated management of alternative strategies is used. The Behaviour Strategy Manager stores a hierarchy of strategies, each of which is guaranteed to achieve different goals under different environment assumptions. The hierarchy is organised in terms of the strength of the assumptions each strategy makes (or the risk that each strategy takes). Higher level strategies make bold assumptions on the environment in order to achieve stronger goals. When the Behaviour Strategy Enactor detects that one of an assumption is violated, it raises an exception to the Behaviour Strategy Manager which then puts in place a behaviour strategy further down in the hierarchy. The new strategy will be based on weaker assumptions compatible with the environment behaviour exhibited up to that point, but at the cost of achieving weaker goals. Indeed, the hierarchy can be used to adapt the functional behaviour of the system through graceful degradation when the assumptions of a higher level model are broken, and through progressive enhancement when those assumptions are satisfied or restored.

In [24] we also present formal results showing that pre-computed behaviour strategies must define a simulation order to guarantee graceful degradation and provide seamless progressive enhancement. We also show how a Goal Model Manager can produce a hierarchy of control problems that will result in hierarchy of simulating strategies using standard controller synthesis algorithms implemented in a Behaviour Problem Solver.

The techniques we present in [15] to compute reconfiguration strategies that must address the construction of a distributed configuration in a decentralised fashion fits into the responsibilities of the Reconfiguration Strategy Solver. In [15] we then show how decentralised self-assembly can be implemented over a gossip protocol.

We have worked extensively on synthesis algorithms for the Behaviour Strategy Solver. These involve constructing strategies with formal winning guarantees against adversarial environments [20], [21], dealing through qualitative reasoning with probabilistic failures in the environment [22] and also with partial goal models [23].

In [39] we present how to update a goal model represented as a logic program in the common knowledge repository by using a probabilistic rule learning approach using feedback from the running system in the form of execution traces. Non-monotonic rule learning based on inductive logic programming finds general rules which explain observations in terms of the conditions under which they occur. The updated models are then used to generate new strategies with a greater chance of success.

## V. Related Work

The last decade has seen a significant build up of the body of work related to engineering self-adaptive systems [44], [45]. Our work builds on this knowledge, emphasising the need to make behaviour and reconfiguration control first-class architectural entities.

The MAPE-K model [46] shows how to structure a control loop in adaptive systems. The four key activities (Monitor, Analysis, Plan and Execute) are performed over a shared data structure that captures the knowledge required for adaptation. The MAPE-K model does not prescribe what knowledge is to be captured nor what aspect of the system is to be controlled. Thus, there is no explicit treatment or distinction between configuration and behaviour adaptation let alone prescribed mechanisms for dealing with coordinated and independent configuration and behaviour adaptation. We design each layer of the reference architecture as a MAPE-K control loop, resulting in a hierarchical control loop structure as in [47].

As discussed in Section II, the architecture proposed builds on those of [2], [3], [46] and others (e.g., [4], [5], [6]). However, to the best of our knowledge existing work does not provide support for both independent and also coordinated structural and behavioural adaptation at the architectural level with exception of [48] and [16] where behaviour and dependency injection strategies are computed separately but are forced to be executed serially.

The MORPH reference architecture is geared towards the use of strategies derived from the field of control engineering referred to as discrete event dynamic system (DEDS) control [49] which naturally fits over the system abstractions used at the architecture level, which is the level we envisage self-adaptation supported by MORPH to operate. DEDS are discrete-state, event-driven systems of which the state evolution depends entirely on the occurrence of discrete events over time. The field builds on, amongst others, supervisory control theory [50], queueing theory [51], and reactive planning [52].

Automated construction of DEDS control strategies have been applied for self-adaptation in many different forms. For instance, in [25] temporal planning is used to produce reconfiguration strategies that do not consider structural constraints and the status of components when applying reconfiguration actions. In [16], an architecture description language (ADL) and a planning-as-model-checking are used to compute and enact reconfiguration strategies. In [7], quantitative analysis and planning are used to compute evolution strategies. In [17], [18], [19] automatic generation of event-based coordination strategies is applied for runtime adaptation of deadlock-free mediators. In [53], a learning technique (the L* algorithm [54]) is applied for automatically generating components behaviour. Note that strategies do not have to be necessarily temporal sequencing of actions or commands. For instance, in [55] reconfiguration strategies used are one-step component parameter changes.

The use of techniques based on control theory for continuous-variable dynamic systems (CVDS) has also significant application to self-adaptation and is also used at the architectural level [56], [57], [10], [11], [58]. Existing techniques of continuous control theory applied to adaptation are single-input single-output (SISO) or multiple-input single-output (MIMO) at best. This differs from discrete event control which tends to be multiple-input multiple-output (MIMO). The controlled variable for these approaches is typically related to an operational parameter of the system configuration (e.g. Thread pool size [59], processor clock speed [9], self-imposed thread sleeps [57], accepted requests per time [60]) thus falling into the category of reconfiguration strategies in our reference architecture. A noteworthy example of such approach is [9], which uses a three tiered scheme that has some parallels with one of the design concerns addressed in this paper. The work in [9] can be understood as a three layered continuous control framework in which the Strategy Enactor implements a linear continuous control strategy. The Strategy Management Layer manages a family of control strategies in which the value of a constant used in the bottom layer is tweaked when the system diverges beyond a threshold, and the Goal Management Layer propagates downwards a family of control strategies computed from scratch based on a higher latency Inference process that analyses the execution log.

Goal modelling notations have been identified as central to self-adaptation [61]. We envision using techniques for modelling adaptation requirements (e.g., [62], [63], [64]) for reasoning about adaptation in the Knowledge Repository. In addition, having a flexible representation of the requirements and the runtime behaviour of the system is also desired. We envisage using approaches such as [65] where an executable modelling language for runtime execution of models (EUREMA) facilitates seamless adaptation.

The need for coordinated control loops to deal with complex self-adaptation scenarios has been identified in [66], [46] amongst others. The MORPH architecture has a hierarchy of MAPE-K loops provided by the Goal Manager, Strategy Manager and Strategy Enactor layers. Furthermore, the bottom two layers are actually implementing two concurrent, yet coordinated, control loops: one for behaviour and the other for reconfiguration control.

The problem of strategy update, required when a strategy in the enactment layer is replaced by a new one, is a crucial part of runtime adaptation and has been studied extensively both in the context of reconfiguration strategies (e.g., [67], [12], [13]) and behaviour strategies [14].

## VI. Conclusions

An architectural approach to self-adaptive systems involves runtime change to both the system configuration and be-haviour. In this paper we propose MORPH , a three-layered reference architecture which separates these aspects, supporting independent and coordinated reconfiguration and behaviour adaptation. This proposal builds on the extensive research work conducted by ourselves and others which provides various techniques and software entities which plug into the architecture; as such MORPH helps to identify and compare the architectural entities investigated and implemented in different systems. There is as yet no comprehensive system covering the full scope of MORPH . Although this is not essential, as demonstrated by the achievements of existing systems, we plan to investigate this further to provide a wider range of reconfiguration and behaviour strategies.

ACKNOWLEDGMENT

This work was partially supported by the following grants: ANPCYT PICT 2012-0724, ANPCYT PICT 2011-1774, ANPCYT PICT 2013-2341, UBACYT 036, UBACYT 0384, CONICET PIP 11220110100596CO, MEALS 295261.